\documentclass[aps,prb,twocolumn,groupedaddress]{revtex4}  % for review and submission
\usepackage{graphicx}  % needed for figures
\usepackage{dcolumn}   % needed for some tables
\usepackage{bm}        % for math
\usepackage{amssymb}   % for math
\usepackage{amsmath}
\usepackage{dsfont}
\usepackage{mathrsfs}
\usepackage{float}

\begin{document}

\newcommand{\bra}[1]{\left\langle#1\right|}
\newcommand{\ket}[1]{\left|#1\right\rangle}
\newcommand{\ev}[1]{\langle#1\rangle}
\newcommand{\cstate}[4]{\ket{\begin{smallmatrix}#1 & #2\\#3 & #4\end{smallmatrix}}}
\newcommand{\Tr}[1]{\mathrm{Tr}\left(#1\right)}

\title{Spectral broadening of optical transitions at tunneling resonances in InAs/GaAs coupled quantum dot pairs}
\author{P. Kumar}\thanks{These authors contributed equally.}
\author{C. Jennings}\thanks{These authors contributed equally.}\email{cjennings2@ucmerced.edu}
\author{M. Scheibner}\email{mscheibner@ucmerced.edu}
\affiliation{School of Natural Sciences, University of California Merced, Merced, California 95343, USA}
\author{A. S. Bracker}
\author{S. G. Carter}
\author{D. Gammon}
\affiliation{Naval Research Laboratory, Washington, DC 20375, USA}
\date{\today}

\begin{abstract}
We report on linewidth analysis of optical transitions in InAs/GaAs coupled quantum dots as a function of bias voltage, temperature, and tunnel coupling strength. A significant line broadening up to 100~$\mu$eV is observed at hole tunneling resonances where the coherent tunnel coupling between spatially direct and indirect exciton states is maximized, corresponding to a phonon-assisted transition rate of $150~\mathrm{ns}^{-1}$ at 20~K. With increasing temperature, the linewidth shows broadening characteristic of single-phonon transitions. The linewidth as a function of tunnel coupling strength tracks the theoretical prediction of linewidth broadening due to phonon-assisted transitions, and is maximized with an energy splitting between the two exciton branches of $0.8-0.9~$meV. This report highlights the linewidth broadening mechanisms and fundamental aspects of the interaction between these systems and the local environment.
\end{abstract}

\maketitle

\section{Introduction}

Vertically stacked coupled quantum dot (CQD) pairs embedded in an electric field effect structure allow for wide-range tuning of atom-like charge and spin states due to an enhanced quantum confined Stark effect (QCSE).\cite{Ramanathan2013,Scheibner2009,Stinaff2006} Optically generated electron-hole pairs can be localized in one of the dots, the electron and hole can be localized in separate dots, or they can be delocalized in both quantum dots by forming a molecular exciton state. The coupling between two dots is quantified by the tunnel coupling strength. The coherent manipulation of exciton states and control of interdot coupling in CQDs offers advantages for CQD-based quantum devices for optical sensing and quantum information processing.\cite{Jennings2019, Kim2011, Loss1998, Scheibner2012, Gao2012, Vora2015, Knill2001, Bayer2001} The pure dephasing, phonon relaxation and charges surrounding the quantum dots are a few major challenges hindering this venture.\cite{Borri2003, Muljarov2005, Bardot2005, Climente2006, Nakaoka2006, Gawarecki2010, Gawarecki2012, Wijesundara2011, Rolon2012, Muller2012, Muller2013, Daniels2013} These phenomena are coupled to the linewidth broadening and line profile, providing details of coupling to the local environment. The pure dephasing expresses the time scale of coherent interactions of charge states with lattice phonons. Other reports highlight charge fluctuation-induced broadening of indirect excitons in CQDs.\cite{Daniels2013,Ha2015,Zhou2013} The linewidth analysis of different charge states and dependence on applied field, including the tunneling resonances where one charge is delocalized, has fundamental research interest and is important to understand the quantum systems for potential quantum computing and sensing applications.

In this paper, we report the detailed analysis of linewidth broadening of direct and indirect excitons and examine the linewidth as a function of electric field near tunneling resonances. We investigate the hypothesis of broadening mechanisms as a function of temperature and tunnel coupling strength. These measurements explore the interaction of the optical transitions in quantum system with the local environment and adjacent charges.

\begin{figure}[H]
\centering
\includegraphics[width=.45\textwidth]{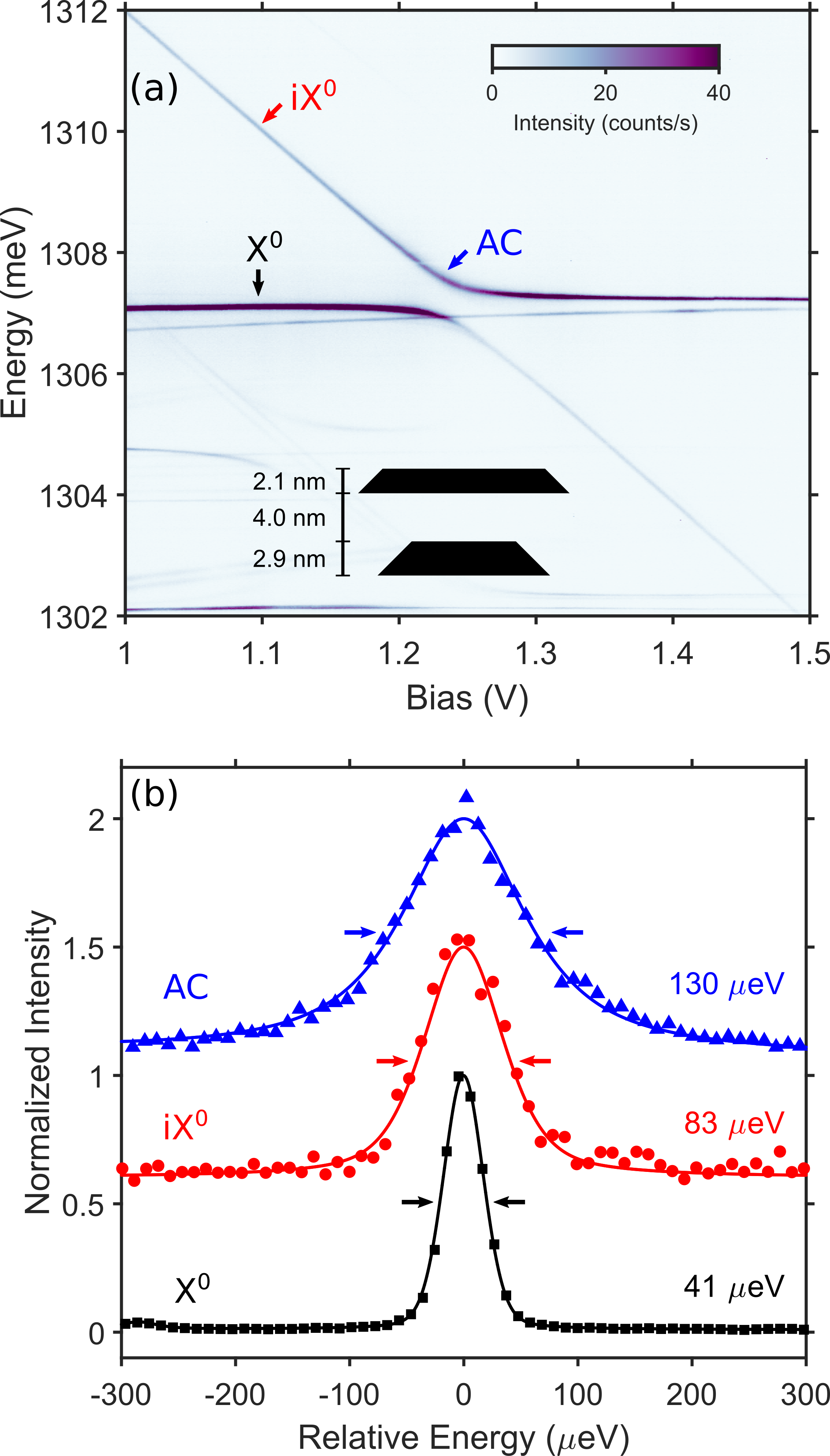}
\caption{(a) Electric field dispersed emission spectra of CQD 1 measured at 20 K near a neutral exciton hole tunneling resonance. Inset shows CQD geometry with QD heights and interdot barrier width. (b) Line profiles of direct ($X^0$, black squares) and indirect ($iX^0$, red circles) optical transitions at $1.1~$V compared with the upper branch tunneling resonance (AC, blue triangles) at $1.24~$V, with Voigt fits (solid lines) and FWHM linewidth values indicated.}
\label{PLmap}
\end{figure}

\section{Experimental Methods and Results}

Molecular beam epitaxy-grown, vertically stacked self-assembled InAs/GaAs quantum dot pairs with 4 nm interdot barrier thickness embedded in a Schottky diode structure are used in this study. The Schottky diode structure allows application of electric field to tune the energy band diagram by shifting the relative energy levels and favors the tunneling of charge carriers between the quantum dots. The details of the fabrication procedure are described elsewhere.\cite{Kerfoot2014} A variable wavelength CW diode laser operating at wavelength $\sim890-950\,$nm and power density $\sim10^{-4}-16\,\mathrm{Wcm}^{-2}$ is used to excite the CQDs quasi-resonantly.  The laser beam is focused on the sample at an angle of 45 degrees to minimize the collection of scattered light. The emission from quantum dot molecules is collected with a 50X magnification microscope objective, dispersed by a triple spectrometer in additive configuration, then subsequently collected using a liquid nitrogen cooled CCD camera. The photoluminescence (PL) energy resolution is limited by the spectrometer response of $30-40~\mu$eV for $30~\mu$m slit opening. A 2600 series Keithley sourcemeter with 6.5-digit resolution is used to apply the electric field along the growth direction of the CQD pair. These measurements are done in a closed cycle helium cooled cryostat at temperatures of $20-80~$K, where $20~$K is the minimum attainable temperature of the cryostat.

\begin{figure}[ht]
\centering
\includegraphics[width=.45\textwidth]{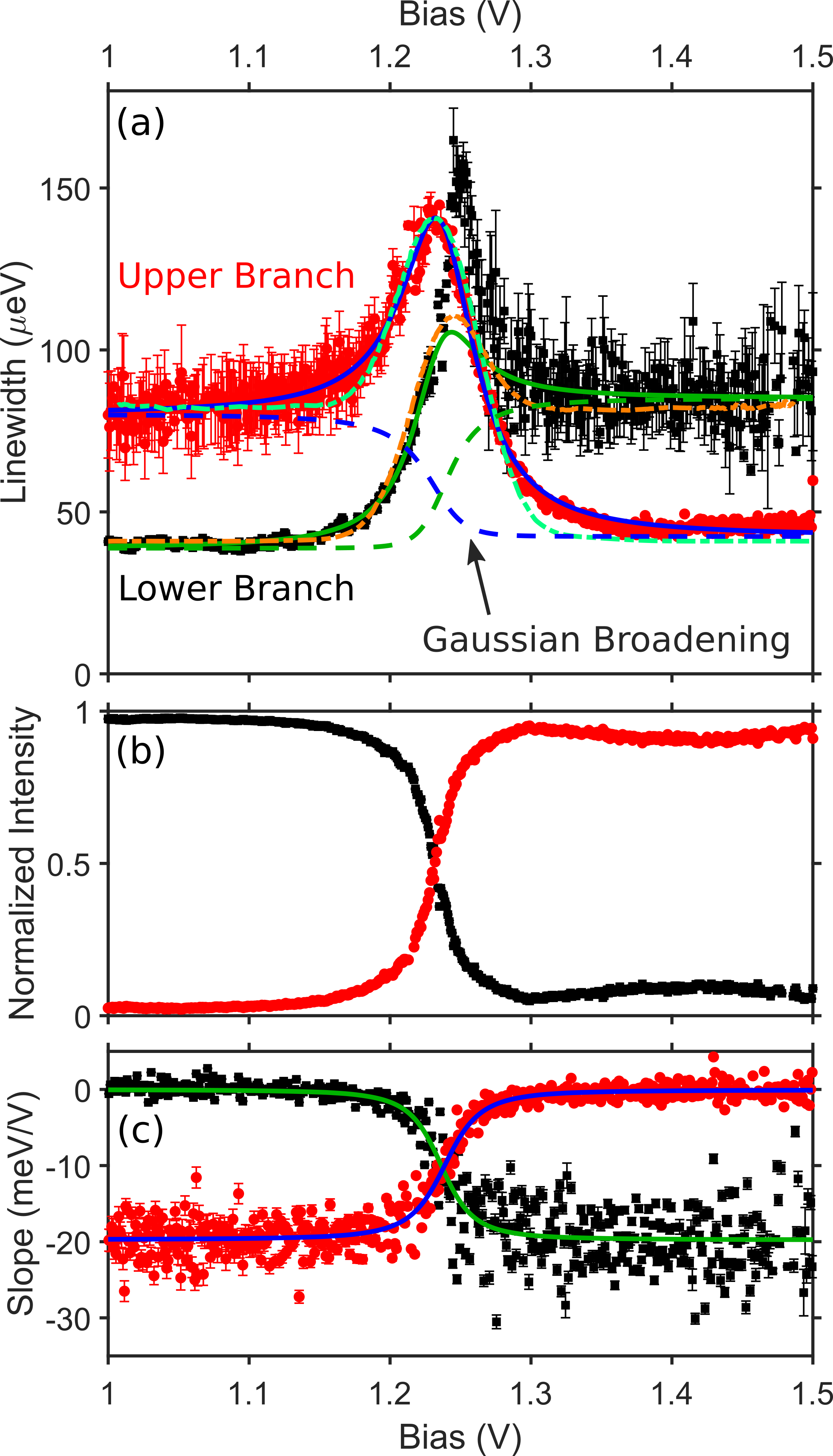}
\caption{(a) Measured linewidth of the lower (black squares) and upper (red circles) branches of the $X^0$ anticrossing in CQD 1 at $20~$K as a function of bias, with fits to Eq.~\eqref{Gfitfunction} (solid lines), the results of numerical simulations (dash-dotted lines), and Gaussian broadening components (dashed lines). (b) Relative intensity of each branch, normalized to the sum of the two intensities at each bias value. (c) Measured bias slope of each branch, with fits to Eq.~\eqref{eigenstateslopes} (solid lines).}
\label{fitdata}
\end{figure}

The characteristic electric field dispersed optical transition energy map for one of the CQDs is shown in Figure~\ref{PLmap}a. This map is generated by fine stepping the applied voltage (in $0.2~$mV increments) along the growth direction of the self-assembled CQD and collecting the corresponding PL. The small separation of 4 nm between the quantum dots allows electron/hole tunneling when electronic levels are brought in resonance by applied electric field skewing the band structure in the intrinsic region of the Schottky-type diode. The PL bias map of the CQD shows multiple optical transitions appearing and disappearing as a function of bias. Every single optical transition is assigned to a charging state based on the spatial location of charge carrier generated. \cite{Scheibner2009,Scheibner2008} Here, we focus our analysis on the spectral broadening of the neutral exciton optical transitions, which generate the two most prominent lines in the electric field dispersed PL spectrum of Fig.~\ref{PLmap}a. The two transitions form an anticrossing (AC) in the center of the image. This anticrossing is a result of a hole level resonance between a direct exciton ($X^0$), with an electron and hole in the bottom dot, and an indirect exciton ($iX^0$), with an electron in the bottom dot and hole in the top dot. The PL emission energy of the direct exciton shows a weak dependence on electric field, while that of the indirect exciton shows a strong electric field dependence. This difference in response to the electric field is a result of the static dipole moment $p={\pm}ed$, defined by the elementary charge $e$ and the spatial separation $d$ of electron and hole. The avoided crossing is the spectral signature of the formation of molecular states, i.e. the symmetric and anti-symmetric mixing of the direct and indirect excitons’ wavefunctions, $\ket{\psi}=\alpha\ket{X}\pm\beta\ket{iX}$.\cite{Doty2009} The resulting exciton state, $\ket{\psi}$, should exhibit properties in between that of the direct and the indirect exciton. For example, at the center of the anticrossing the Stark shift is the average of the Stark shift observed for the direct and indirect excitons. Likewise, the radiative lifetime at the center of the anticrossing can be expected to be the arithmetic average of the lifetimes of both exciton states. Consequently, if we were to measure the linewidth of the exciton transition as we follow one of the anticrossing branches through the anticrossing region, i.e. from the direct exciton to the indirect exciton, we expect the linewidths to gradually and monotonically decrease in the absence of nonradiative broadening mechanisms.

The line profiles for three different exciton states $X^0$, $iX^0$, and tunneling resonance are shown in Figure~\ref{PLmap}b. The solid lines are pseudo-Voigt fits to the experimental data, evaluated as a linear combination of Lorentzian and Gaussian lineshapes. The linewidth of the direct exciton corresponds to the resolution limit of our experimental setup, $41.4\pm0.1~\mu$eV. In contrast, we find that the PL linewidth of the indirect exciton is $83\pm5~\mu$eV, while the linewidth at the upper branch of the anticrossing is $130\pm3~\mu$eV. In resonant measurements, resolution limited by the laser linewidth, such as described by Czarnocki et al.,\cite{Czarnocki2016} we have been able to show that the actual transition linewidth of the direct exciton is on the order of a few $\mu$eV, consistent with the typical radiative lifetimes for InAs/GaAs QDs.\cite{Bardot2005} For the indirect exciton one would expect a much-reduced linewidth, due to the reduced overlap of the electron and hole wavefunctions. That the indirect exciton transitions exhibit the opposite, a larger linewidth than the direct exciton transitions, has been attributed to charge fluctuations near the CQDs and the larger static dipole moment.\cite{Ramanathan2013,Zhou2013,Ha2015} Regardless of this inverted behavior of the linewidths, we expect a gradual and monotonic change of the exciton transition linewidth as we follow one of the branches through the anticrossing.

In contrast to the expected behavior, we find a non-monotonic change of the PL linewidth. Towards the center of the anticrossing the linewidth increases to values significantly above that of the indirect exciton. In the example shown in Fig.~\ref{fitdata}a, the linewidth of the upper branch broadens at the tunneling resonance to $139\pm1~\mu$eV compared to $42.5\pm0.6~\mu$eV in the limit of the direct exciton transition and $80.6\pm0.6~\mu$eV in the limit of the indirect exciton transition, with similar values for the lower branch. We investigated more than 20 molecules and observed linewidth broadening up to $\sim300~\mu$eV at the tunneling resonance. Theoretical work by Daniels et al. suggests that the linewidth broadening at the anticrossing is the result of enhanced phonon coupling.\cite{Daniels2013} They find that at the tunneling resonances where the two involved exciton states come closest in energy to each other, transition rates between the two branches assisted by the emission or absorption of phonons are enhanced.

The relative intensities of the upper and lower exciton branches are shown in Fig.~\ref{fitdata}b. The intensity of each branch is equal near the tunneling resonance, where the wavefunction overlap is maximized. The indirect exciton becomes significantly weaker in intensity away from the tunneling resonance, leading to increased uncertainty of linewidth fit values. The slope (change in exciton peak energy as a function of applied bias) of the upper and lower branches is shown in Fig.~\ref{fitdata}c, and follows the predicted dependence of Eq.~\ref{eigenstateslopes} with equal slopes at the tunneling resonance.

\begin{figure}[ht]
\centering
\includegraphics[width=.45\textwidth]{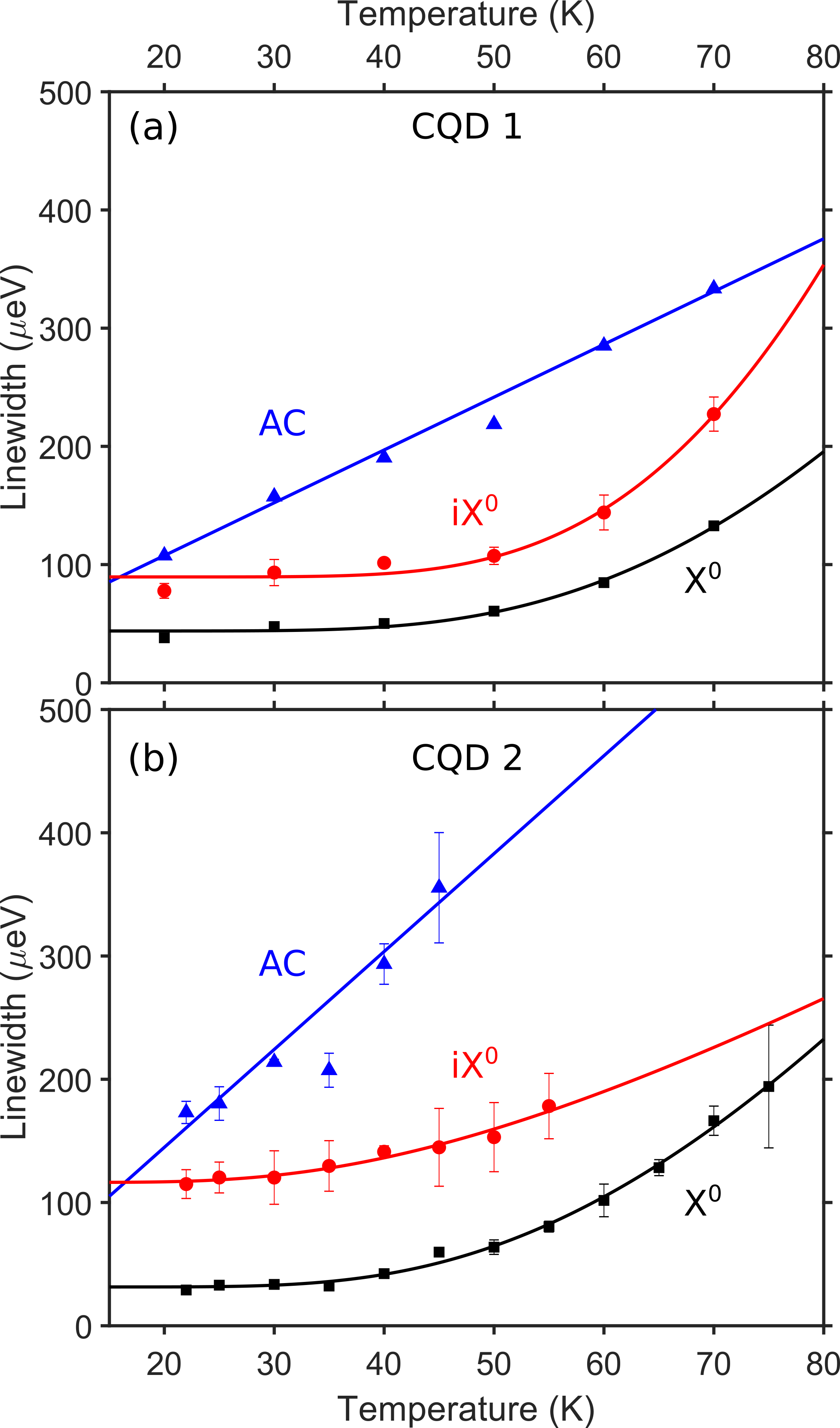}
\caption{Measured temperature-dependent ZPL linewidths for (a) CQD 1 and (b) CQD 2 at and away from the center of the anticrossing, with linear and Bose broadening fits (solid lines).}
\label{tempseries}
\end{figure}

\begin{figure}[ht]
\centering
\includegraphics[width=.45\textwidth]{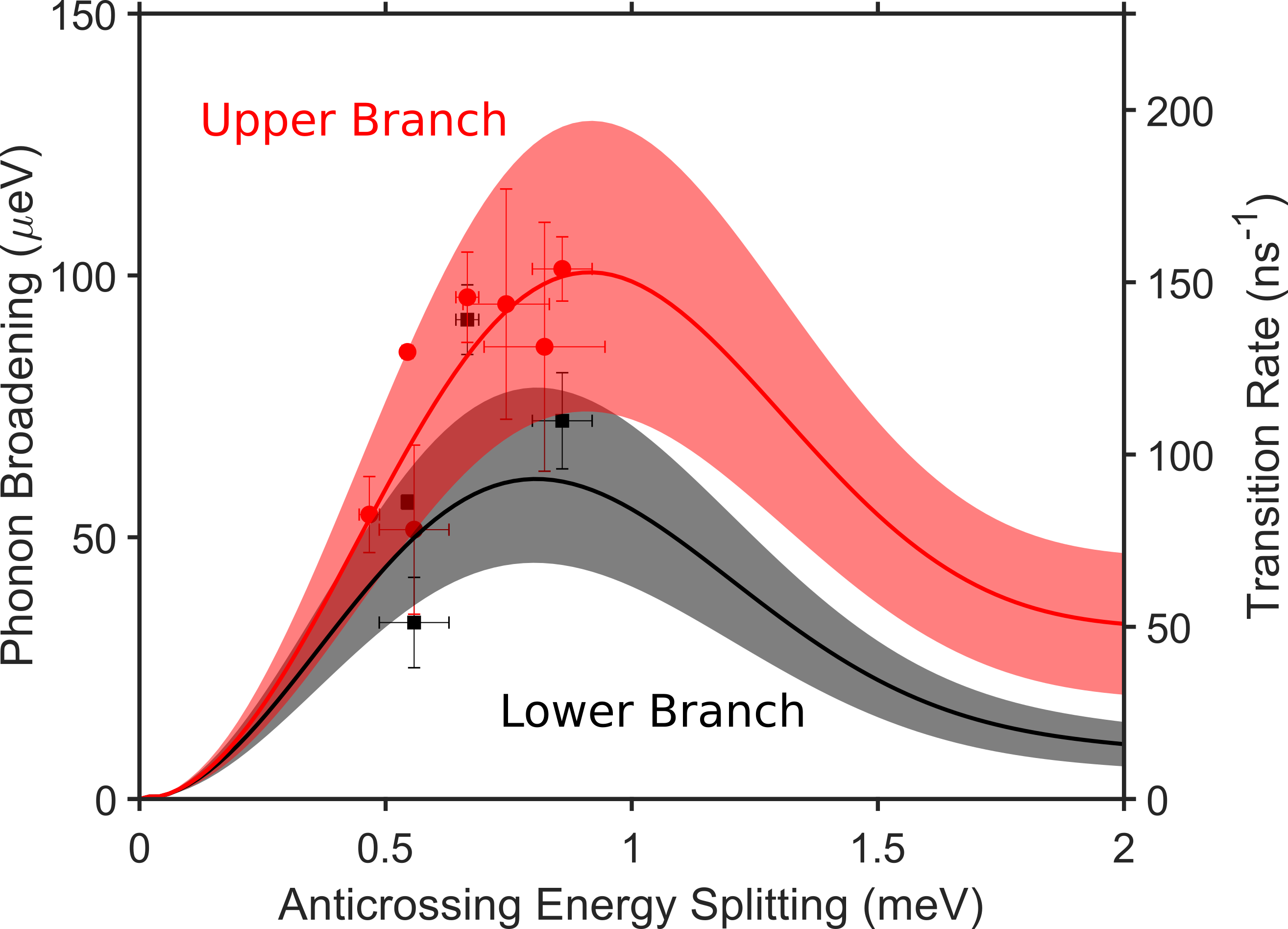}
\caption{Phonon broadening and corresponding transition rates of the lower (black squares) and upper (red circles) branches of the $X^0$ anticrossing for 7 CQDs at $20~$K as obtained from fitting to Eq.~\eqref{Gfitfunction}, compared with numerical simulations (solid lines) of perfectly aligned QDs at a fixed interdot barrier width of $4~$nm. Shaded regions show simulations for a range of coupling parameters matching observations.}
\label{ACsplittingseries}
\end{figure}

The temperature dependence of the PL linewidth is shown in Fig.~\ref{tempseries} for two CQDs on the same sample, with the theoretical dependence given by Eq's.~\eqref{transitionrate} and \eqref{energylinewidth}. At low temperatures, the linewidth is determined by the one-phonon transition rate between the lowest two eigenstates, with the temperature dependence entering through the phonon mode population $n_B(T,\omega_{21})$ at the transition frequency. The energy splitting $\hbar{\omega_{21}}$ at the anticrossing is significantly smaller than the thermal energy $k_B T$ in these measurements, leading to the observed linear broadening for the upper branch
\begin{equation}\label{linearbroadening}
\frac{\partial \Gamma_2}{\partial T} \approx \frac{\pi}{\hbar} \frac{k_B}{\hbar \omega_{21}} J_{12}(\omega_{21}).
\end{equation}
The slope of this linear broadening, measured as $7.9\pm2.1~{\mu}$eV/K for CQD 1 and $4.5\pm0.4~{\mu}$eV/K for CQD 2, is therefore proportional to the interdot phonon coupling strength through the spectral density $J_{12}(\omega_{21})$ at the transition frequency. The difference in broadening slope and phonon coupling strength between CQD 1 and CQD 2 can be explained by a variation of lateral alignment between QDs, as discussed in Section~\ref{disc}. This explanation is supported by the observation of different tunnel coupling strengths $230\pm10~\mu$eV and $540\pm20~\mu$eV of the first two excited states of the neutral exciton in CQD 1, corresponding to coupling between the ground state of the bottom QD and the first two excited states of the top QD with $p$-like orbitals $8.6\pm0.1~$meV and $14.1\pm0.1~$meV above the ground state, respectively.\cite{Scheibner2008} The different energies of the $p_x$-like and $p_y$-like excited states indicate an elongated top QD, while the larger tunnel coupling strength of the second excited state indicates a lateral misalignment between QDs along the shorter axis of the top QD. The measurements were limited to a temperature $\geq 20~$K due to the closed cycle cryostat system used. We expect that the linewidth at the anticrossing would approach a constant value at lower temperatures between $5-15~$K where the thermal energy decreases below the energy splitting.

The effect of energy splitting between exciton branches on the phonon-induced linewidth broadening for 7 CQDs is shown in Fig.~\ref{ACsplittingseries}. The value of phonon broadening $\Gamma^{ph}_{1/2}$ for each branch is obtained by fitting the bias-dependent linewidth to Eq.~\ref{Gfitfunction} to remove the effects of Gaussian broadening due to charge fluctuations and spectrometer resolution. The results are compared with numerical simulations of perfectly aligned QDs with an interdot barrier width of $4~$nm, predicting a maximum broadening of $100~\mu$eV at $0.9~$meV for the upper branch and $60~\mu$eV at $0.8~$meV for the lower branch. This corresponds to a maximum transition rate of $150~\mathrm{ns}^{-1}$ ($90~\mathrm{ns}^{-1}$) for phonon emission (absorption). The experimental data appears to follow the simulated curve with variations of up to $\pm28\%$ from predictions using average phonon coupling strength. The inferred transition rates are comparable to measured\cite{Muller2012,Muller2013} and calculated\cite{Climente2006,Gawarecki2010,Gawarecki2012,Daniels2013} electron/hole interdot relaxation rates in the $0.5-1.0~$meV energy range, though substantially lower rates have also been observed.\cite{Nakaoka2006,Wijesundara2011,Rolon2012}

\section{Theoretical Model}

To describe the optical transitions of a tunnel-coupled quantum dot pair (CQD) in an electric field-effect diode structure near a resonant tunneling anticrossing, we follow Refs.~\onlinecite{Daniels2013} and \onlinecite{Gawarecki2010} by starting with a two-band effective mass model to describe bound charges in the conduction and heavy-hole valence bands. This gives single-particle wavefunctions $\Psi^{\alpha}_{i{\sigma}}(\vec{r})=\psi^{\alpha}_i(\vec{r})u^{\alpha}_{\sigma}(\vec{r})$ with band index $\alpha=\{e,h\}$, QD location $i=\{B,T\}$, spin state $\sigma$, and lattice-periodic Bloch wavefunctions $u^{\alpha}_{\sigma}(\vec{r})$. The confinement model and resulting envelope wavefunctions $\psi^{\alpha}_i(\vec{r})$ are detailed in Appendix~\ref{wavefunctions}.

Near the ground state hole tunneling resonance of the neutral exciton state, the spatially direct and indirect excitons expressed in the localized basis as $\ket{X}=\ket{e_B}\ket{h_B}$ and $\ket{iX}=\ket{e_B}\ket{h_T}$, respectively, are coupled to form new eigenstates
\begin{equation}\label{eigenstates}
\begin{aligned}
\ket{1} &= a_{11}\ket{X} + a_{12}\ket{iX} \\
\ket{2} &= a_{21}\ket{X} + a_{22}\ket{iX}.
\end{aligned}
\end{equation}
The coefficients $a_{ij}$ are found by diagonalizing the Hamiltonian matrix\cite{Bayer2001}
\begin{equation}\label{tunneling hamiltonian}
H_X =
\begin{pmatrix}
E_X(U) & -t_h \\
-t_h & E_{iX}(U)
\end{pmatrix},
\end{equation}
where $t_h$ is the hole tunnel coupling energy and $E_{X(iX)}(U)$ is the experimentally determined energy of state $\ket{X}(\ket{iX})$ as a function of bias voltage $U$ applied to the diode. The eigenstate energies
\begin{equation}\label{eigenstateenergies}
\begin{aligned}
E_{1/2}=&\frac{E_X(U)+E_{iX}(U)}{2} \\
&\mp\sqrt{\left(\frac{E_X(U)-E_{iX}(U)}{2}\right)^2+t_h^2}
\end{aligned}
\end{equation}
form an avoided crossing, or anticrossing, with the minimum energy difference at resonance given by ${\Delta}E_{\mathrm{min}}=2t_h$. Using the linear approximation of Stark shift near an anticrossing centered at $U_{AC}$, the eigenstate energies are given by $E_X(U)=E_0$ and $E_{iX}(U)=E_0-p(U-U_{AC})$, leading to the bias-dependent slopes
\begin{equation}\label{eigenstateslopes}
\frac{\partial E_{1/2}}{\partial U} = -\frac{p}{2} \left( 1 \pm \frac{U-U_{AC}}{\sqrt{(2t_h/p)^2+(U-U_{AC})^2}} \right)
\end{equation}
for each eigenstate. 

Coupling between single bound charges and lattice phonons can be described using the general Hamiltonian\cite{Climente2006,Gawarecki2010,Gawarecki2012,Daniels2013}
\begin{equation}\label{phonon hamiltonian}
\begin{aligned}
H_{e-ph} = & \sum_{s,\vec{q}} (b_{s,\vec{q}}+b^{\dagger}_{s,-\vec{q}}) \\
&\times \left[ \sum_{ij} c^{\dagger}_i c_j F^e_{s,ij}(\vec{q}) - \sum_{kl} d^{\dagger}_k d_l F^h_{s,kl} (\vec{q}) \right],
\end{aligned}
\end{equation}
with creation (annihilation) operators $b^{\dagger}_{s,{\vec{q}}}$ ($b_{s,{\vec{q}}}$) for phonon modes with polarization $s=\{\mathrm{LA},\mathrm{TA_1},\mathrm{TA_2}\}$ and wave vector $\vec{q}$, $c^{\dagger}_i$ ($c_i$) for electrons in state $\ket{i}$, and $d^{\dagger}_k$ ($d_k$) for holes in state $\ket{k}$. The phonon coupling constants are expanded into bulk and localized contributions as $F^{\alpha}_{s,ij}(\vec{q})=g^{\alpha}_s(\vec{q})\mathcal{F}^{\alpha}_{ij}(\vec{q})$, with bulk coupling matrix elements $g^{\alpha}_s(\vec{q})$ depending on phonon mode and coupling mechanism and geometric form factors
\begin{equation}\label{formfactor}
\mathcal{F}^{\alpha}_{ij}(\vec{q})=\bra{\alpha_i}e^{i\vec{q}\cdot\vec{r}}\ket{\alpha_j}=\int \psi^{{\alpha}*}_i(\vec{r})e^{i\vec{q}\cdot\vec{r}}\psi^{\alpha}_j(\vec{r})\,d^3\vec{r}
\end{equation}
describing overlap of the envelope wavefunctions of involved states modulated by the phonon mode phase.

Since we are interested in transitions between the two lowest-energy neutral exciton states near a tunneling resonance, the relevant energy differences are less than 15 meV, so coupling to optical phonons at energies of 30-40 meV is neglected. The relevant phonon coupling mechanisms which contribute to the bulk matrix element $g^{\alpha}_s(\vec{q})$ therefore include deformation potential (DP) coupling to LA phonons, given by
\begin{equation}\label{DPcoupling}
g^{e/h(DP)}_{LA}(\vec{q})=\sqrt{\frac{{\hbar}q}{2{\rho}Vc_{LA}}}a_{c/v},
\end{equation}
and piezoelectric (PE) coupling to LA and TA phonons, given by
\begin{equation}\label{PEcoupling}
g^{\alpha(PE)}_s(\vec{q})=-i\sqrt{\frac{\hbar}{2{\rho}Vc_sq}}\frac{d_Pe}{\epsilon_0\epsilon_r}M_s(\hat{q}).
\end{equation}
In equations \eqref{DPcoupling} and \eqref{PEcoupling}, $\rho$ is the mass density of the crystal, $V$ is the crystal volume used for normalization of phonon modes (cancels out after summation over wave vectors), $c_s$ is the propagation velocity of phonon mode $s$, $a_{c/v}$ is the deformation potential of the conduction/valence band, $d_P$ is the piezoelectric constant of the crystal, $\epsilon_0\epsilon_r$ is the electric permittivity of the crystal, and the directional dependence $M_s(\hat{q})$ of the PE coupling is detailed in Appendix~\ref{simmethods}. Note that these bulk coupling matrix elements assume a constant value of each material parameter, without taking into account variations in composition due to the CQD structure. Previous studies therefore assume that these parameters are determined entirely by the GaAs barrier material, or by assuming a uniform effective composition.\cite{Gawarecki2010,Daniels2013}

The single-particle phonon coupling Hamiltonian given by Eq.~\eqref{phonon hamiltonian} can be transformed to the diagonalized exciton basis as
\begin{equation}\label{exciton phonon hamiltonian}
H_{X-ph} = \sum_{nm} \sum_{s,\vec{q}} F^X_{s,nm}(\vec{q})\ket{n}\bra{m}(b_{s,\vec{q}}+b^{\dagger}_{s,-\vec{q}}),
\end{equation}
where the exciton-phonon coupling constants $F^X_{s,nm}$ are obtained by projecting Eq.~\eqref{phonon hamiltonian} onto the diagonalized eigenstates. Transitions between states $\ket{1}$ and $\ket{2}$ necessarily involve hole tunneling, such that electron-phonon coupling does not contribute. Pure dephasing processes $\ket{n}\rightarrow\ket{n}$ with no population transfer, describing phonon-assisted optical transitions, can occur by electron- or hole-phonon coupling. Taking these properties into account, the exciton-phonon coupling constants are given in terms of the localized single-particle coupling constants as
\begin{equation}\label{exciton phonon coupling}
\begin{aligned}
F^X_{s,11} =& F^e_{s,BB} - a_{11}^2 F^h_{s,BB} - a_{12}^2 F^h_{s,TT} - 2 a_{11} a_{12} F^h_{s,BT} \\
F^X_{s,22} =& F^e_{s,BB} - a_{21}^2 F^h_{s,BB} - a_{22}^2 F^h_{s,TT} - 2 a_{21} a_{22} F^h_{s,BT} \\
F^X_{s,12} =& - a_{11} a_{21} F^h_{s,BB} - a_{12} a_{22} F^h_{s,TT} \\
& - (a_{11} a_{22} + a_{12} a_{21}) F^h_{s,BT}.
\end{aligned}
\end{equation}

\begin{figure}[ht]
\centering
\includegraphics[width=.45\textwidth]{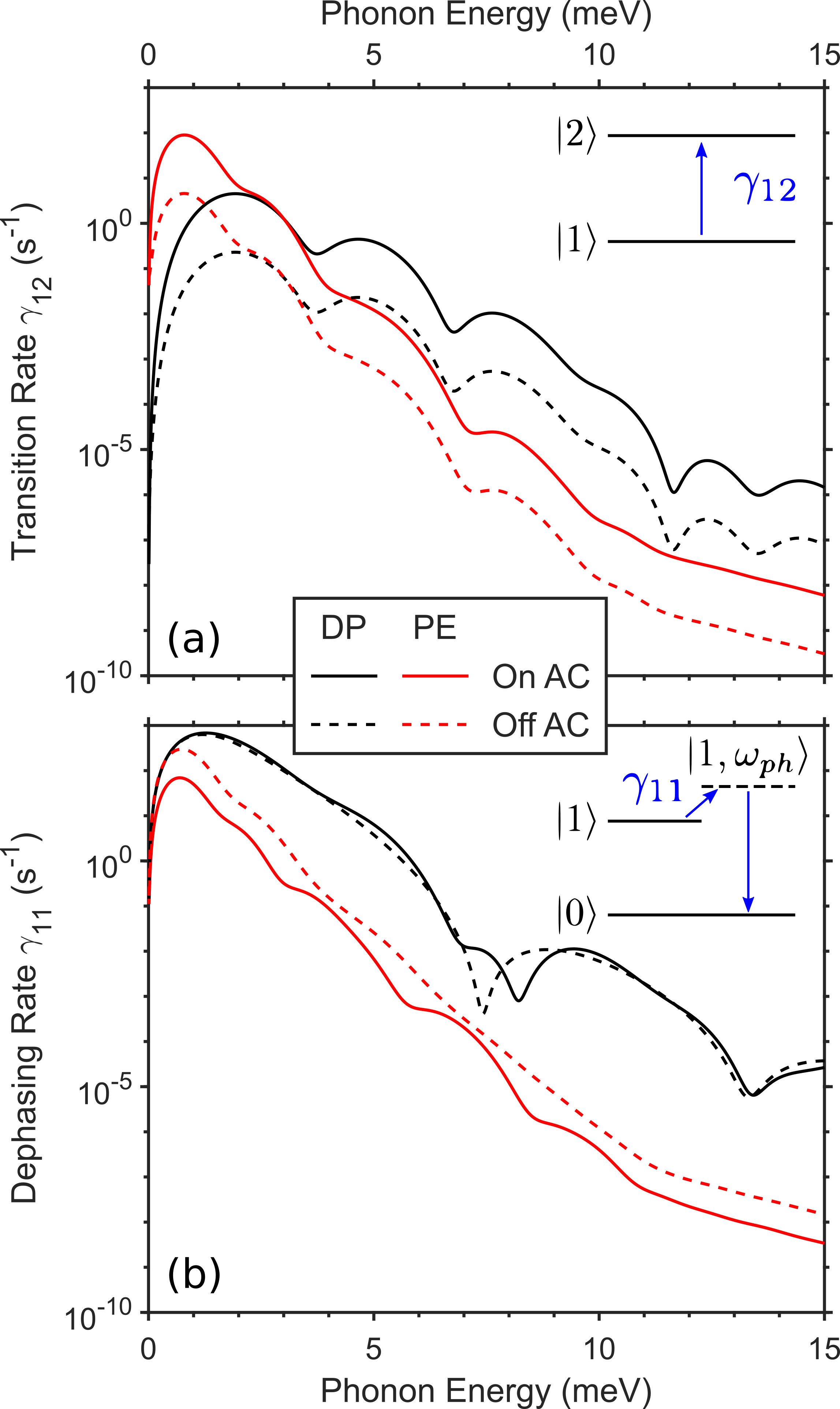}
\caption{Simulated (a) transition rate and (b) pure dephasing rate calculated for CQD 1 at the center of the anticrossing (at 1.24 V, solid lines) and away from the anticrossing (at 1.10 V, dashed lines), with contributions from deformation potential coupling (black lines) and piezoelectric coupling (red lines). Insets depict the selected transition or dephasing process.}
\label{relaxationdephasing}
\end{figure}

The rate of phonon-assisted tunneling transitions from state $\ket{n}$ to state $\ket{m}$ due to first-order coupling is given by Fermi's golden rule as
\begin{equation}\label{transitionrate}
\gamma_{nm}=\frac{2\pi}{\hbar^2}\left[n_B(T,|\omega_{nm}|)+\Theta(\omega_{nm})\right]J_{nm}(|\omega_{nm}|),
\end{equation}
where the phonon spectral density
\begin{equation}\label{spectraldensity}
J_{nm}(\omega)=\sum_{s,\vec{q}} \left| F^X_{s,nm}(\vec{q}) \right|^2 \delta(\omega-\omega_{s,\vec{q}})
\end{equation}
measures the coupling to phonon modes at the transition frequency $\omega_{nm}=(E_n-E_m)/\hbar$ to ensure energy conservation, the temperature-dependent phonon mode population is given by the Bose distribution $n_B(T,\omega)=(e^{\hbar\omega/{k_B}T}-1)^{-1}$, and the step function $\Theta(\omega_{nm})=0\,(1)$ for phonon absorption (emission).

The general expression for the energy linewidth of each exciton state
\begin{equation}\label{energylinewidth}
\Gamma_n(\omega)=2\hbar \tilde{\gamma}_n(\omega)=\hbar\sum_{m{\neq}n}\gamma_{nm}+\hbar\gamma_{nn}(\omega)
\end{equation}
contains contributions from real transitions to other states as well as virtual single-state transitions associated with phonon-assisted optical absorption, resulting in acoustic phonon sidebands around the Lorentzian zero-phonon line (ZPL) and pure dephasing.\cite{Besombes2001,Favero2003,Borri2003,Borri2005} Here we consider only the lowest-energy direct and indirect states, neglecting excited states, which are expected to be $10-20~$meV higher in energy and have a negligibly small tunneling rate. The frequency-dependent pure dephasing rate is calculated similarly to the tunneling rate, with the phonon energy determined by the detuning $\Delta\omega_n=\omega-\omega_n$ of the optical frequency from resonance:
\begin{equation}\label{puredephasing}
\gamma_{nn}(\omega)=\frac{2\pi}{\hbar^2}\left[n_B(T,\Delta\omega_n)+\Theta(\Delta\omega_n)\right]J_{nn}(\Delta\omega_n).
\end{equation}

Experimentally, the emission spectrum is detected using a spectrometer with a finite resolution. As a result, the detected spectrum is convolved with the typically Gaussian spectrometer response function of width $\Gamma_{spect}$ ($30-40~\mu$eV in these experiments), leading to a Voigt ZPL profile with phonon sidebands. In the presence of a fluctuating electric field due to many charged lattice defects near the CQD, an additional Gaussian broadening is present, with a width $\Gamma_{fluct,1/2}={\Delta}U_{fluct}|{\partial}E_{1/2}/{\partial}U|$ proportional to the bias slope of the transition energy given in Eq.~\ref{eigenstateslopes}. With both of these broadening mechanisms, the combined Gaussian ZPL broadening is given by
\begin{equation}\label{gaussianbroadening}
\Gamma_{g,1/2} = \sqrt{ (\Gamma_{fluct, 1/2})^2 + (\Gamma_{spect})^2 }.
\end{equation}

\section{Discussion}\label{disc}

The non-monotonic bias dependence of linewidth in Fig.~\ref{fitdata}a, together with the additional temperature-dependent broadening in Fig.~\ref{tempseries}, indicate a significant enhancement of phonon-assisted transition rates between eigenstates at tunneling resonances. The bias-dependent ZPL linewidth can be fit to the predicted form of Gaussian broadening in Eq.~\ref{gaussianbroadening} with an additional phonon-induced broadening with a Lorentzian shape, resulting in the function
\begin{widetext}
\begin{equation}\label{Gfitfunction}
\Gamma_{1/2}(U) = \sqrt{\left(\frac{p{\Delta}U_{fluct}}{2}\right)^2\left(1\pm\frac{U-U_{AC}}{\sqrt{(2t_h/p)^2+(U-U_{AC})^2}}\right)^2+(\Gamma_{spect})^2} + \frac{\Gamma^{ph}_{1/2}}{1+\left(\frac{U-U_{AC}}{2t_h/p}\right)^2}
\end{equation}
\end{widetext}
for each branch, with fit parameters $U_{AC}$, $t_h$, and $p$ describing the position and shape of the anticrossing energy levels and ${\Delta}U_{fluct}$, $\Gamma_{spect}$, and $\Gamma^{ph}_{1/2}$ describing the strength of broadening due to charge fluctuations, spectrometer resolution, and phonon-assisted transitions, respectively.

\begin{figure}[ht]
\centering
\includegraphics[width=.45\textwidth]{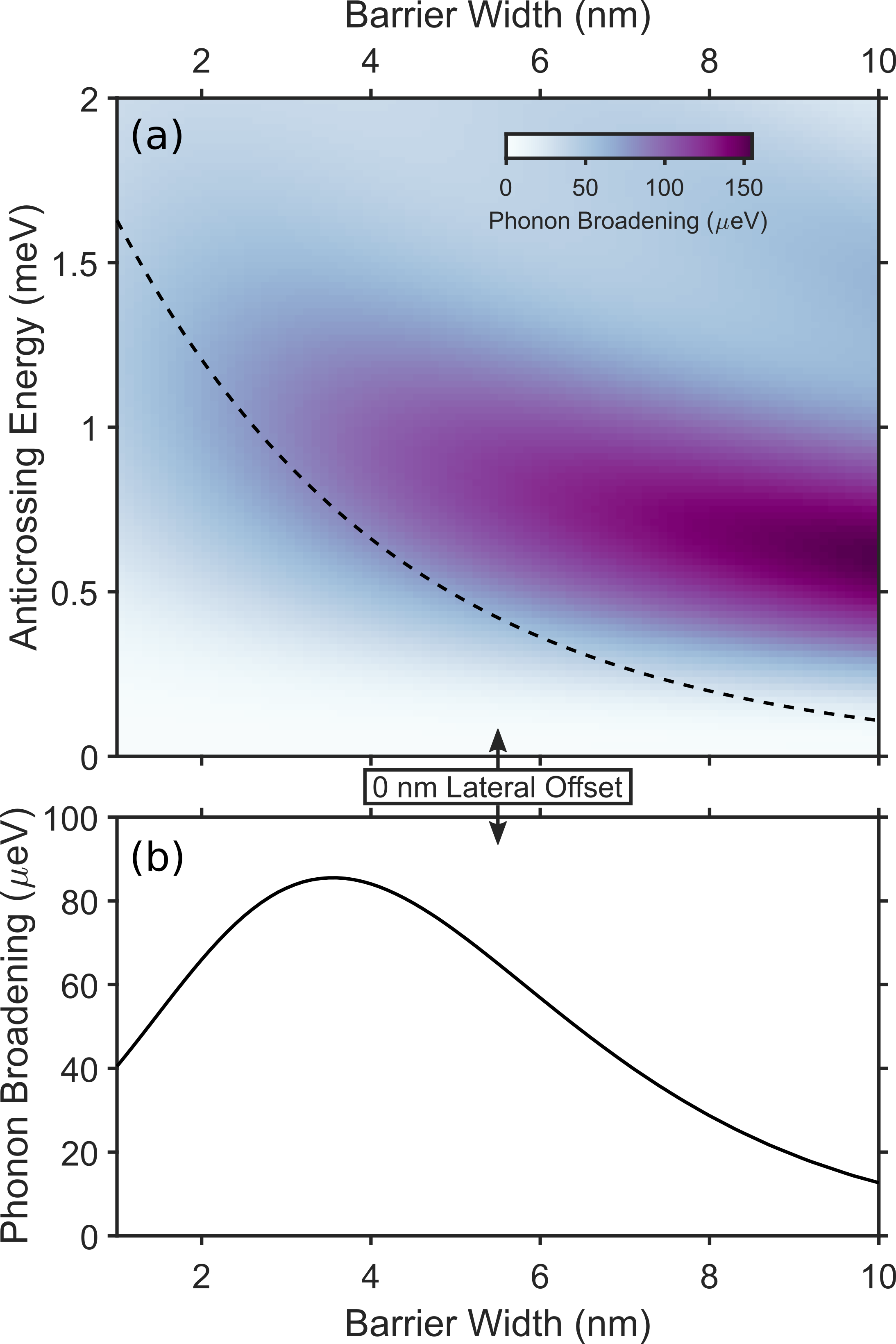}
\caption{(a) Simulated upper branch phonon broadening at 20~K and 0~nm lateral offset, as a function of anticrossing energy and interdot barrier width. Dashed line shows anticrossing energy values expected from previous experimental observations.\cite{Bracker2006} (b) Simulated phonon broadening for expected anticrossing energies as a function of barrier width.}
\label{barrierdependence}
\end{figure}

\begin{figure}[ht]
\centering
\includegraphics[width=.45\textwidth]{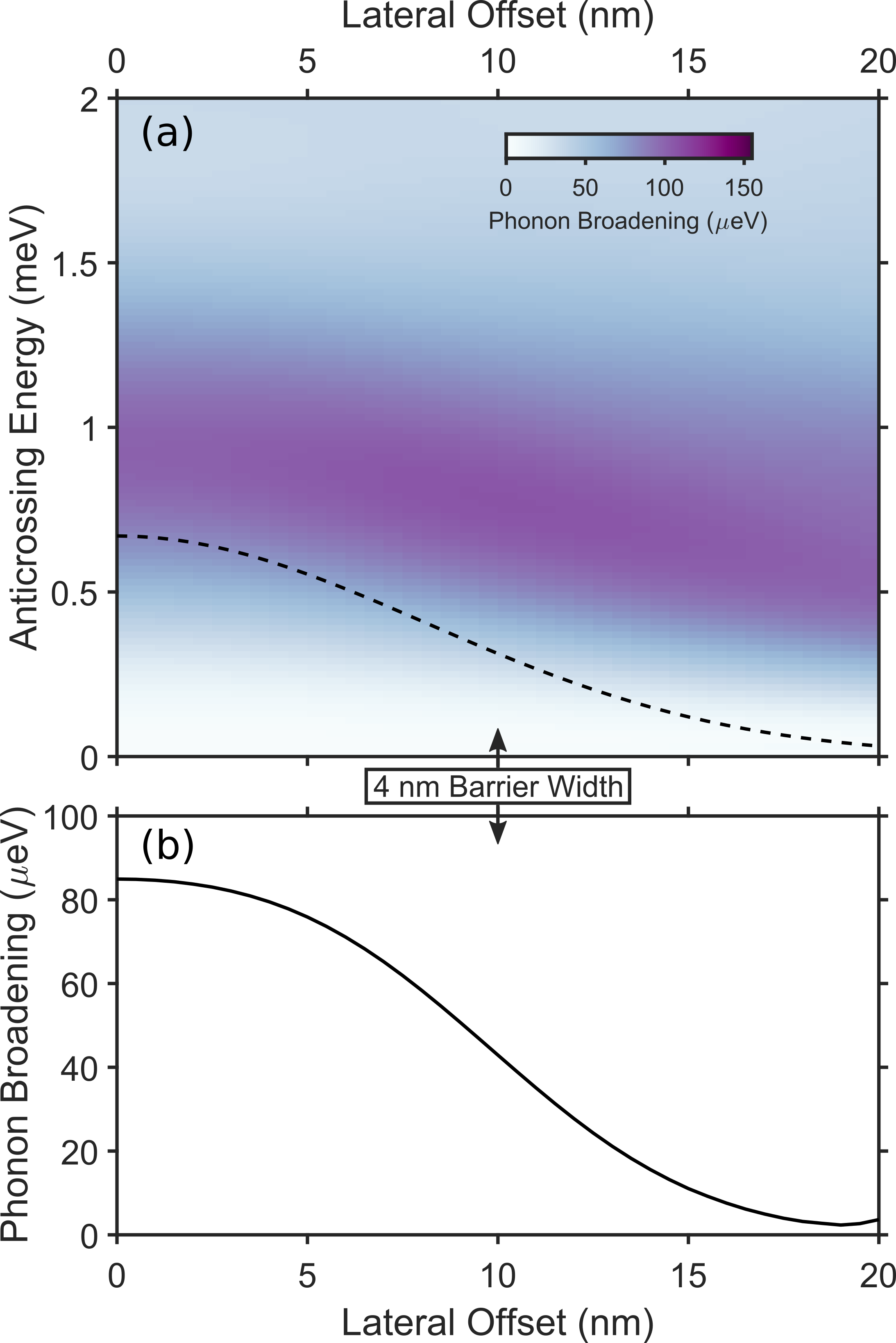}
\caption{(a) Simulated upper branch phonon broadening at 20~K and 4~nm barrier width, as a function of anticrossing energy and lateral offset between QD centers. Dashed line shows anticrossing energy values expected from previous experimental observations.\cite{Bracker2006} (b) Simulated phonon broadening for expected anticrossing energies as a function of lateral offset.}
\label{offsetdependence}
\end{figure}

The theory predicts an asymmetry in peak linewidths of the upper and lower branches, with the upper branch being more broad due to a faster phonon emission process compared to phonon absorption, resulting in a shorter lifetime for state $\ket{2}$. Fits to observed spectra appear to show an additional broadening of the lower branch just past the center of the anticrossing, to a level higher than the peak linewidth of the upper branch. However, the region with increased fit linewidth of the lower branch corresponds to where a second faintly visible peak merges with it. This peak appears to be due to a weakly-allowed recombination from the dark exciton spin state due to spin-orbit coupling, with an exchange splitting of $225\pm14~\mu$eV far from the anticrossing.\cite{Daniels2013,Doty2010}

Fig.~\ref{relaxationdephasing} shows the calculated phonon-assisted transition and pure dephasing rates both on and off the anticrossing resonance for each coupling mechanism at 20~K, as a function of phonon energy. Transitions between eigenstates are dominated by piezoelectric coupling at low energies, with a maximum at $0.8~$meV for phonon absorption from the lower branch. The linewidth broadening effect is therefore predicted to be strongest for CQDs with an anticrossing splitting energy of $0.8-0.9~$meV. The sideband-producing pure dephasing process is dominated by deformation potential coupling, with a maximum at 1.2~meV. The experimental spectra should therefore give a measure of piezoelectric coupling strength through the ZPL linewidth at the anticrossing and deformation potential coupling through the intensity and distribution of acoustic phonon sidebands, which are more prominent away from the anticrossing where the ZPL is narrower. The oscillatory decay of transition and dephasing rates as a function of phonon energy is a known feature of CQDs, arising from resonances in the phonon coupling form factor (Eq.~\ref{formfactor}) between phonon wavelength and QD separation.\cite{Climente2006,Gawarecki2010,Wijesundara2011,Rolon2012} The complex oscillation pattern and its bias dependence are due to the combination of single-particle coupling components between localized basis states, as calculated from Eq.~\ref{exciton phonon coupling}. The phonon coupling strength $F^X_{s,12}(\vec{q})$, primarily due to the piezoelectric interaction, was initially too high when calculated using the material parameters for GaAs listed in Table~\ref{parameters}. This was reduced to match observed peak anticrossing linewidths by using effective $\mathrm{In}_x\mathrm{Ga}_{1-x}\mathrm{As}$ composition values of $x=32\pm12\%$ when calculating phonon coupling constants, with material parameters varying linearly between GaAs ($x=0$) and InAs ($x=1$). The deformation potential parameters $a_c$ and $a_v$ were both increased by a factor between 1.27 and 2.22 relative to the values listed in Table~\ref{parameters} to match observed sideband intensities between $0.6\%$ and $1.4\%$ of ZPL intensity, since literature values of these parameters are highly inconsistent. The resulting values of InAs composition and phonon coupling strength match our experimental observations within the two-band effective mass model and are consistent with reports of In migration during QD growth,\cite{Liu2000,Kegel2001,Lenz2002,Jovanov2012} though inclusion of coupling with light-hole valence bands may significantly modify these values.\cite{Climente2008,Gawarecki2012} The coupling strength could also be modified more weakly through the form factor $\mathcal{F}^{\alpha}_{ij}(\vec{q})$ by changing the QD charge confinement and single-particle wavefunctions.

The variations in phonon coupling strength can potentially be explained by differences in CQD geometry throughout the sample, with simulated dependence on interdot barrier width shown in Fig.~\ref{barrierdependence} and on lateral misalignment in Fig.~\ref{offsetdependence}. While the interdot barrier width is expected to be quite uniform throughout each sample, variances in CQD alignment have been observed and could significantly reduce the phonon coupling depending on the lateral confinement within each QD.\cite{Doty2010} Since the value of tunnel coupling and anticrossing energy is proportional to wavefunction overlap between localized states, we calibrate the value of anticrossing energy expected in each case using previous measurements on a series of CQD samples grown similarly with different interdot barrier widths to obtain the curves in the lower plots.\cite{Bracker2006} The simulations predict maximum phonon broadening for interdot barrier widths near 4~nm, and a decrease in phonon broadening with lateral QD misalignment.

While the data and simulations presented in this report focus on the hole tunneling resonance of the neutral exciton state, we expect that the enhancement of phonon coupling at tunneling resonances is a more general effect which can apply to different charge states as well. The geometric phonon coupling form factor is increased by the formation of delocalized eigenstates, which occurs at any tunneling resonance regardless of the configuration of resident charges. The bulk PE coupling constant (Eq.~\ref{PEcoupling}) is equal for electrons and holes, so the effect can occur regardless of which charge carrier is tunneling. The only remaining requirement for strong phonon coupling enhancement is that the AC splitting energy lies near the maximum of the phonon spectral density for PE coupling, a condition which depends on the size and confinement potential of the QDs. Electron tunneling ACs typically have a much larger energy splitting due to their lower effective mass, inhibiting this effect since PE coupling is strongly weighted towards lower phonon energies.\cite{Bracker2006} The effect might be observed with electron tunneling by reducing AC splitting energy through proper band engineering of the interdot barrier\cite{Liu2011} or by working with excited-state ACs.\cite{Scheibner2008,Muller2012,Muller2013} Initial observations indicate a similar level of phonon broadening at hole tunneling resonances in positive trion and neutral biexciton transitions, though the presence of additional optically active spin states makes the fitting procedure more complicated and the results less reliable.

\section{Conclusion}

To conclude, we have measured the linewidths of direct exciton, indirect exciton and tunneling resonance states for CQDs. We find that pure dephasing, phonon relaxation and charge fluctuations in the CQDs can explain the observed linewidth broadening. The existence of phonon transitions between the molecular-like excitons in the system causes the linewidth to broaden beyond the charge fluctuation-induced broadening of the indirect exciton state. The transition of linewidths from direct to indirect exciton state is non-monotonic near tunneling resonances and phonon-induced broadening up to 100~$\mu$eV is reported at 20~K, corresponding to phonon-assisted transition rates up to $150~\mathrm{ns}^{-1}$. These measurements are in good agreement with theoretical calculations of linewidth broadening at tunneling resonances including phonon-assisted transitions due to PE and DP coupling.

\begin{acknowledgments}
We acknowledge funding from the Defense Threat Reduction Agency (Grant No. HDTRA1-15-1-0011). A.S.B., S.G.C., and D.G. acknowledge the support of Office of Naval Research.
\end{acknowledgments}

\appendix

\section{Single-Particle Wavefunctions}\label{wavefunctions}
The slowly-varying envelope wavefunctions $\psi^{\alpha}_i(\vec{r})$ are solutions to the Schr\"{o}dinger equation
\begin{equation}\label{schrodinger}
\left[\frac{-\hbar^2}{2m_{\alpha}}\nabla^2 + V_{\alpha}(\vec{r})\right] \psi^{\alpha}_i(\vec{r}) = E^{\alpha}_i \psi^{\alpha}_i(\vec{r})
\end{equation}
with effective mass $m_{\alpha}$, confinement potential $V_{\alpha}(\vec{r})$ and single-particle confinement energies $E^{\alpha}_i$.

Eq.~\eqref{schrodinger} can be simplified by modeling the confinement potential of each quantum dot (QD) using the cylindrically symmetric function
\begin{equation}\label{confinementpotential}
V_{\alpha}(r,\phi,z)=\mathcal{E}_{\alpha}\Theta\left(|z|-\frac{h}{2}\right) + \frac{1}{2}m_{\alpha}{\omega}^2_{\alpha}r^2
\end{equation}
to describe finite well confinement in the vertical direction due to band-edge offsets $\mathcal{E}_{\alpha}$ of the heterostructure and harmonic oscillator confinement in the lateral direction, where $\Theta(x)$ is the Heaviside step function, $h$ is the height of the QD, and the angular frequency $\omega_{\alpha}$ of the lateral harmonic oscillator is set by the experimentally determined spacing ${\hbar}{\omega}_{\alpha}$ between the ground and first excited states. Eq.~\eqref{schrodinger} is then solved using separation of variables to give envelope wavefunctions of single-particle localized ground states as
\begin{equation}\label{wavefunction}
\psi^{\alpha}_1(\vec{r})=Ae^{-m_{\alpha}\omega_{\alpha}r^2/2\hbar}Z_1(z),
\end{equation}
where $A$ is a normalization constant defined such that $\int |\psi^{\alpha}_i(\vec{r})|^2 \, d^3\vec{r} = 1$ and the $z$-component of the wavefunction is expressed as the piecewise function
\begin{equation}\label{Zwavefunction}
\begin{aligned}
Z_n(z)=
\begin{cases}
cos(k_nz) & \text{if } |z|\;{\leq}\;h/2 \\
cos(k_nh/2)e^{-\kappa_n(|z|-h/2)} & \text{if } |z|>h/2,
\end{cases}
\end{aligned}
\end{equation}
with wavenumber $k_n$ determined as the $n$-th solution to the transcendental equation
\begin{equation}\label{transcendentalequation}
\tan^2\left(\frac{k_nh}{2}\right)=\frac{m_{\alpha,InAs}}{m_{\alpha,GaAs}}\left(\frac{k_0^2}{k_n^2}-1\right)
\end{equation}
due to the boundary conditions for continuity of the wavefunction and its first derivative,
\begin{equation}\label{wavefunctioncontinuity}
\kappa_n^2=k_0^2-\left(\frac{m_{\alpha,\mathrm{GaAs}}}{m_{\alpha,\mathrm{InAs}}}\right)k_n^2,
\end{equation}
and $k_0^2=2m_{\alpha,\mathrm{InAs}}\mathcal{E}_{\alpha}/\hbar^2$. In a coordinate system with the origin set at the center between the two QDs and assuming no lateral misalignment such that cylindrical symmetry is preserved, wavefunctions for particles localized in each dot are found from Eqs.~\eqref{wavefunction} and \eqref{Zwavefunction} as $\psi^{\alpha}_{B/T}(r,\phi,z)=\psi^{\alpha}_1(r,\phi,z \mp d/2)$, with the substitution $h \mapsto h_{B/T}$ to account for the different height of each QD and center-to-center QD separation $d$.

We note that Refs.~\onlinecite{Daniels2013} and \onlinecite{Gawarecki2010} use a more detailed geometrical model of the CQD, treating the confinement potential as a pair of lens-shaped finite wells. They use an adiabatic separation of variables technique to solve the 1-D Schr\"{o}dinger equation with finite double well potential in the vertical direction at each radial distance and use the resulting radius-dependent confinement energy as an additional potential term in the radial Schr\"{o}dinger equation. Finally, the Ritz variational method is applied to approximate the eigenstates as linear combinations of the obtained vertical and radial wavefunctions which minimize the total energy. Ref.~\onlinecite{Gawarecki2010} additionally uses a continuum elasticity model to calculate the strain distribution and obtain spatially-dependent components of the anisotropic effective mass tensor, both of which are used as inputs to the eigenstate calculations.

\section{Simulation Methods}\label{simmethods}
For numerical simulations, each integral is converted to a sum over a grid of values with a sufficient number of grid points to achieve satisfactory convergence. Since the system is assumed to have cylindrical symmetry, single-particle localized ground state wavefunctions are calculated using Eq.~\eqref{wavefunction} and represented in cylindrical coordinates $\vec{r}=(r,\phi,z)$ as a product of $r$- and $z$-dependent components $\psi^{\alpha}_{B/T}(\vec{r})=R^{\alpha}_0(r)Z^{\alpha}_1(z{\mp}d/2)$. Both $r$ and $z$ values are represented as 100-point grids, covering $30~$nm in the $r$ direction and $20~$nm in the $z$ direction. For simulations varying lateral offset between QDs (Fig.~\ref{offsetdependence}), wavefunctions are represented in Cartesian coordinates with a 25-point grid in each dimension since cylindrical symmetry is broken. Due to the delta function in Eq.~\eqref{spectraldensity} which enforces energy conservation, it is most convenient to express phonon wavevectors in terms of energy in spherical coordinates $\vec{q}_s=(E/{\hbar c_s},\phi,\theta)$. Both angular coordinates are represented as 200-point grids covering a full $4\pi$ solid angle, with the azimuthal coordinate $\phi$ from $0$ to $2\pi$ and the polar coordinate $\theta$ from $0$ to $\pi$.

The directional dependence of the PE coupling is given in terms of the phonon mode polarization vectors $\hat{e}_{s,\vec{q}}$ as
\begin{equation}\label{Mgeneral}
M_s(\hat{q})=2\left[\hat{q}_x(\hat{e}_{s,\vec{q}})_y\hat{q}_z+\hat{q}_y(\hat{e}_{s,\vec{q}})_z\hat{q}_x+\hat{q}_z(\hat{e}_{s,\vec{q}})_x\hat{q}_y\right].
\end{equation}
Using the phonon mode polarization vectors
\begin{equation}\label{phononpolarization}
\begin{aligned}
\hat{e}_{LA,\vec{q}}\equiv\hat{q}&=(\cos{\phi}\sin{\theta},\sin{\phi}\sin{\theta},\cos{\theta})\\
\hat{e}_{TA_1,\vec{q}}&=(-\sin{\phi},\cos{\phi},0)\\
\hat{e}_{TA_2,\vec{q}}&=(\cos{\phi}\cos{\theta},\sin{\phi}\cos{\theta},-\sin{\theta}),
\end{aligned}
\end{equation}
equation \eqref{Mgeneral} for each phonon mode becomes
\begin{equation}\label{Mcomponents}
\begin{aligned}
M_{LA}(\hat{q})&=\frac{3}{2}\sin(2\phi)\sin(2\theta)\sin{\theta}\\
M_{TA_1}(\hat{q})&=\cos(2\phi)\sin(2\theta)\\
M_{TA_2}(\hat{q})&=\sin(2\phi)\sin{\theta}(3\cos^2{\theta}-1).
\end{aligned}
\end{equation}
Due to the cylindrical symmetry, $M_s(\hat{q})$ is averaged over the azimuthal coordinate as $\bar{M}_s(\theta)=\left(\int_0^{2\pi}M_s(\phi,\theta)^2\,d\phi/{2\pi}\right)^{1/2}$ to obtain
\begin{equation}\label{Mcylindrical}
\begin{aligned}
\bar{M}_{LA}(\theta) &= \sqrt{\frac{9}{8}} \sin(2\theta) \sin{\theta} \\
\bar{M}_{TA_1}(\theta) &= \frac{1}{\sqrt{2}} \sin(2\theta) \\
\bar{M}_{TA_2}(\theta) &= \frac{1}{\sqrt{2}} \sin{\theta} (3\cos^2{\theta}-1).
\end{aligned}
\end{equation}

Evaluation of the geometric form factors $\mathcal{F}^{\alpha}_{ij}(\vec{q})$ defined in Eq.~\eqref{formfactor} involves integration over a three-dimensional grid of spatial coordinates for each value of the phonon wave vector on a separate three-dimensional grid, thereby constituting a major bottleneck in numerical calculations. Ref.~\onlinecite{Gawarecki2010} uses the cylindrical symmetry of the envelope wavefunctions to simplify these integrals by separating variables and evaluating the angular integral in terms of $m$'th-order Bessel functions of the first kind $\mathcal{J}_m(a)$. For the separable ground-state wavefunctions defined in Eq.~\eqref{wavefunction} and phonon wave vectors defined in cylindrical coordinates as $\vec{q}=(q_r,\phi,q_z)$, this expression becomes
\begin{equation}\label{formfactorcyl}
\begin{aligned}
\mathcal{F}^{\alpha}_{ij}(\vec{q})=&2\pi\int_0^{\infty} e^{-m_{\alpha}\omega_{\alpha}r^2/\hbar} \mathcal{J}_0(q_{r}r) \,r dr \, \\
& \times \int_{-\infty}^{\infty} Z_i(z) e^{i{q_z}z} Z_j(z)\, dz .
\end{aligned}
\end{equation}
The form factor is then expressed in spherical coordinates using the transformations $q_r=q\sin{\theta}$ and $q_z=q\cos{\theta}$.

Finally, single-particle coupling constants $F^{\alpha}_{s,ij}(\vec{q})$, calculated using the obtained form factors and bulk coupling constants given by Eqs.~\eqref{DPcoupling} and \eqref{PEcoupling}, are represented for each particle $\alpha=\{e,h\}$ and set of QD locations $\{i,j\}=\{B,T\}$ as a function of phonon mode $s$, phonon energy $E$, and polar angle $\theta$. The summation over phonon modes is represented in spherical coordinates as an integral over wave vectors with a fixed magnitude:
\begin{equation}\label{spectraldensityintegral}
\begin{aligned}
J_{nm}(\omega)=&\frac{V}{(2\pi)^3}\sum_s\frac{\omega^2}{c_s^3} \\
&\times\int_0^{\pi}\int_0^{2\pi}\left|F^X_{s,nm}\left(\omega/c_s,\phi,\theta\right)\right|^2\,\sin{\theta}\,d{\phi}d\theta,
\end{aligned}
\end{equation}
where the dispersion relation $E={\hbar}{\omega}={\hbar}{c_s}q$ is used to relate wave vector magnitude to the mode-dependent group velocity $c_s$, and the mode volume $V$ cancels with the corresponding factor in the bulk coupling constants $g^{\alpha}_s(\vec{q})$.

Ref.~\onlinecite{Daniels2013} uses a single-particle Green's function description of linear susceptibility within the electric dipole and rotating wave approximations to obtain an expression for the optical absorption spectrum as a sum of Lorentzian contributions
\begin{equation}\label{absorption}
I_{abs}(\omega)\propto \sum_n |M_n|^2 \frac{\tilde{\gamma}_n(\omega)}{\left(\omega-E_n/\hbar\right)^2+\tilde{\gamma}_n(\omega)^2},
\end{equation}
where $M_n$ is the optical dipole matrix element of transition $\ket{0}\rightarrow\ket{n}$. These matrix elements are expressed in terms of electron-hole wavefunction overlaps
\begin{equation}\label{ehoverlap}
M_{ij} = \int \psi^{e*}_i(\vec{r}) \psi^h_j(\vec{r}) \,d^3\vec{r}
\end{equation}
of localized exciton states, giving
\begin{equation}\label{overlapmixing}
\begin{aligned}
M_1 &= a_{11}^2 M_{BB} + a_{12}^2 M_{BT} \\
M_2 &= a_{21}^2 M_{BB} + a_{22}^2 M_{BT}.
\end{aligned}
\end{equation}
The optical emission spectrum is also calculated similarly to the absorption spectrum, with only the pure dephasing rates modified by changing the sign of detuning terms $\Delta\omega_n$ to $-\Delta\omega_n$ to reflect the reversal of phonon absorption and emission processes.

At each value of the bias voltage $U$, the tunneling Hamiltonian given by Eq.~\eqref{tunneling hamiltonian} is diagonalized to obtain the eigenstate coefficients $a_{ij}$. These are used to obtain the phonon coupling constants $F^X_{nm}$ and optical dipole matrix elements $M_n$ in the eigenstate basis. The phonon spectral density can then be calculated using Eq.~\eqref{spectraldensityintegral}, allowing calculation of the optical transition rates and absorption and emission spectra using Eq.~\eqref{absorption}. As a final step, Gaussian convolutions are applied to the optical transition spectra in the energy and bias directions to reproduce broadening due to the spectrometer response and local charge fluctuations, respectively. The values of material and structural parameters used in the simulations are listed in Table~\ref{parameters}, except where otherwise noted.

\section{Parameter Values}
\begin{table}[H]
\begin{tabular}{lccc}
\hline \hline
 & & GaAs & InAs \\
\hline
\multicolumn{2}{c}{\underline{Material Parameters}} & & \\ 
Electron effective mass ($m_0$) \cite{Barker2000} & $m_e$ & 0.059 & 0.042 \\ 
Hole effective mass ($m_0$) \cite{Barker2000} & $m_h$ & 0.37 & 0.34 \\ 
Conduction band edge (eV) \cite{Barker2000} & $E_c$ & 1.518 & 1.057 \\ 
Valence band edge (eV) \cite{Barker2000} & $E_v$ & 0 & 0.192 \\ 
CB deformation potential (eV) \cite{Gawarecki2010} & $a_c$ & \multicolumn{2}{c}{-9.3} \\ 
VB deformation potential (eV) \cite{Gawarecki2010} & $a_v$ & \multicolumn{2}{c}{-0.7} \\
Piezoelectric constant ($\mathrm{C}/\mathrm{m}^2$) \cite{Gawarecki2010} & $d_P$ & 0.16 & 0.045 \\
Relative dielectric constant \cite{Gawarecki2010} & $\epsilon_r$ & 12.9 & 15.15 \\
Crystal density ($\mathrm{kg}/\mathrm{m}^3$) \cite{Gawarecki2010} & $\rho$ & 5300 & 5670 \\
LA phonon velocity (m/s) \cite{Gawarecki2010} & $c_{LA}$ & \multicolumn{2}{c}{5150} \\
TA phonon velocity (m/s) \cite{Gawarecki2010} & $c_{TA}$ & \multicolumn{2}{c}{2800} \\
\multicolumn{2}{c}{\underline{Quantum Dot Parameters}} & & \\
Bottom QD height (nm) \cite{Scheibner2008} & $h_B$ & \multicolumn{2}{c}{2.9} \\
Top QD height (nm) \cite{Scheibner2008} & $h_T$ & \multicolumn{2}{c}{2.1} \\
QD center separation (nm) & $d$ & \multicolumn{2}{c}{6.5} \\
$e^-$ excited state spacing (meV) & $\hbar{\omega}_e$ & \multicolumn{2}{c}{100} \\
$h^+$ excited state spacing (meV) \cite{Scheibner2008} & $\hbar{\omega}_h$ & \multicolumn{2}{c}{21.2} \\
$h^+$ tunnel coupling (${\mu}\mathrm{eV}$) & $t_h$ & \multicolumn{2}{c}{330.5} \\
Exciton intensity ratio & $I_X/I_{iX}$ & \multicolumn{2}{c}{17.09} \\
Bias fluctuation width (mV) & ${\Delta}U_{fluct}$ & \multicolumn{2}{c}{3.58} \\
\multicolumn{2}{c}{\underline{Experiment Parameters}} & & \\
Temperature (K) & $T$ & \multicolumn{2}{c}{20} \\
Spectrometer resolution (${\mu}\mathrm{eV}$) & $\Gamma_{spect}$ & \multicolumn{2}{c}{37.0} \\
\hline \hline
\end{tabular}
\caption{Numerical values of physical parameters used in all simulations, except where otherwise noted. Values are taken from references where specified.}
\label{parameters}
\end{table}

\bibliographystyle{apsrev}

\end{document}